# Production of exotic electromagnetic bound systems in ultra-peripheral heavy ion collisions with two-photon processes

Gongming Yu[1], Zhongxia Zhao[2*], Yanbing Cai[3], Quangui Gao[4], Qiang Hu[5], Haitao Yang[2]

[1]Department of Physics, Kunming University, Kunming 650214, China
[2]College of Science, Zhaotong University, Zhaotong 657000, China
[3] Guizhou Key Laboratory in Physics and Related Areas, and Guizhou Key Laboratory of Big Data Statistic Analysis, Guizhou University of Finance and Economics, Guiyang 550025, China
[4]Department of Physics, Yuxi Normal University, Yuxi 653100, China
[5]Institute of Modern Physics, Chinese Academy of Sciences, Lanzhou 730000, China

We calculate the production of exotic electromagnetic bound systems, which ia an consisting of a ($l^+l^-$) bound state system such as positronium ($e^+e^-$), dimuonium ($\mu^+\mu^-$), and ditauonium ($\tau^+\tau^-$), by photon-photon interaction with the equivalent photon approximation in ultra-peripheral heavy ion collisions considering the nuclear form factor. The numerical results demonstrate that the experimental study of positronium, dimuonium, and ditauonium in ultra-peripheral collisions is feasible at Relativistic Heavy Ion Collider (RHIC) and Large Hadron Collider (LHC) energies.

## I. INTRODUCTION

The study of exotic electromagnetic bound systems denoted here as ($l^+l^-$) and their properties is of theoretical and experimental interest. The observation of positronium ($e^+e^-$) [1] and the clarification of the leptonic nature of muons lead to a natural question that whether there are dimuonium ($\mu^+\mu^-$) [2-4] and ditauonium ($\tau^+\tau^-$) [5]. The bound systems ($l^+l^-$) such as positronium, dimuonium, and ditauonium, are the most compact pure quantum electrodynamics (QED) systems, that can be used to test fundamental laws like the CPT theorem [6]. Many methods have been proposed to create the ($l^+l^-$) bound systems, such as the photoproduction [7-10] and electroproduction [11,12] on nuclei and atoms. The exotic electromagnetic bound systems (positronium, dimuonium, and ditauonium) can also be produced in relativistic heavy ion collisions through a pure electromagnetic process [13], where the mechanism is the interaction of two photons emitted from the nuclei. Recently, the possibility of ($l^+l^-$) production at modern electron-positron colliders is investigated [14]. Although, the ($l^+l^-$) bound systems has also been subject to extensive theoretical investigations [15-25], dimuonium and ditauonium has not yet been experimentally discovered.

In the present work, we consider the purely electromagnetic production process,
$$A_1 + A_2 \rightarrow A_1 + A_2 + (l^+l^-),$$
where $A_i$ denotes the relativistic heavy nuclei with charge numbers $Z_i$, and ($l^+l^-$) is the exotic electromagnetic bound systems denoted as positronium ($e^+e^-$), dimuonium ($\mu^+\mu^-$), and ditauonium ($\tau^+\tau^-$). Since the nuclei do not change during the ($l^+l^-$) production process, they emit the photons coherently or semi-coherently. Because the constant $Z_i\alpha \approx 0.6$ ( $\alpha \approx 1/137$ ) is the electromagnetic coupling constant) for Au and Pb is large, two-photon interactions may be accompanied by additional electromagnetic interactions, where the photon comes from the electromagnetic field of nucleus. This leads to a very large flux of equivalent photons available for the ($l^+l^-$) production. Here, we calculate the cross sections, rapidity and transverse momentum distributions for two-photon production process of a selection of exotic electromagnetic bound systems ($l^+l^-$). Because the cross section for two-photon interaction process scales as $Z^4$, we focus on heavy nuclei uranium-uranium collisions with a center-of-mass energy of $\sqrt{s_{NN}} = 193$ GeV per nucleon, gold-gold collisions with a center-of-mass energy of 200GeV per nucleon at the Relativistic Heavy Ion Collider (RHIC), and lead-lead collisions with an center-of-mass energy of 2.76TeV and 5.5TeV per nucleon at the Large Hadron Collider (LHC).

In this paper, we report on a feasibility study of the coherently and semi-coherently two-photon production process for exotic electromagnetic bound systems denoted as ($l^+l^-$) at RHIC and LHC. In Sec.II we present the coherently and semi-coherently two-photon production process for positronium (e$^+$e$^-$), dimuonium ($\mu^+\mu^-$), and ditauonium ($\tau^+\tau^-$) at LHC energies. The numerical results for positronium, dimuonium, and ditauonium production in U+U and Au+Au collisions at RHIC energies, and Pb+Pb collisions at LHC energies are plotted in Sec.III. Finally, the conclusion is given in Sec.IV.

## II. GENERAL FORMALISM

According to the equivalent photons approximation, the cross section of exotic electromagnetic bound systems ($l^+l^-$) produced by two-photon process can be factorized into an elementary cross section for $\gamma\gamma \to$ ($l^+l^-$) and a $\gamma\gamma$ luminosity. The cross section to produce a final state with mass $W$ can be written as

$$d\sigma = \hat{\sigma}_{\gamma\gamma \to (l^+l^-)}(W)dN_1(\omega_1,q_1^2)dN_2(\omega_2,q_2^2)$$
$$= d\omega_1 d\omega_2 \hat{\sigma}_{\gamma\gamma \to (l^+l^-)}(W)\frac{dN_1(\omega_1,q_1^2)}{d\omega_1}\frac{dN_2(\omega_2,q_2^2)}{d\omega_2}, \quad (1)$$

where the energies for the photons emitted from the nucleus are $\omega_{1,2} = \frac{W}{2}\exp(\pm y)$, with $W^2 = 4\omega_1\omega_2$, and the transformations $d\omega_1 d\omega_2 = (W/2)dWdy$ can be performed. The total cross section of the real photoproduction process $\hat{\sigma}_{\gamma\gamma \to (l^+l^-)}(W)$ for the exotic electromagnetic bound systems (positronium, dimuonium, and ditauonium) production can be easily obtained using the known decay width of the $n^1S_0$ state and summing over all states [26-31]

$$\hat{\sigma}_{\gamma\gamma \to (l^+l^-)}(W) = 8\pi^2(2J+1)\frac{\Gamma_{(l^+l^-)\to\gamma\gamma}}{M}(W^2-M^2), \quad (2)$$

here $J$ and $M$ are the spin and mass of the produced exotic electromagnetic bound state (positronium, dimuonium, and ditauonium), respectively.

Since only the probability density for $S$-state at the origin does not vanish, which implies that $|\Psi_{ns}(0)|^2 = \frac{\alpha^3 M}{8\pi n^3}$. Consequently, the decay width $\Gamma_{(l^+l^-)\to\gamma\gamma}$ for lowest ($l^+l^-$) state $n^1S_0$ can be expresses as [5, 30]

$$\Gamma_{(l^+l^-)\to\gamma\gamma} = \frac{\alpha^5 M}{2n^3}, \quad (3)$$

where α is the electromagnetic coupling constant.

The γγ luminosity is given by convolution of the equivalent photon spectra from the two relativistic nuclei with Z times the electric charge moving with a relativistic factor $\gamma \gg 1$ with respect to some observer develops an equally strong magnetic-field component. In the equivalent photon approximation, it resembles a beam of real photons where the photon spectrum function can be written as [32–35]

$$\frac{dN(\omega,q^2)}{d\omega} = \frac{Z^2\alpha}{\pi\omega} \int d^2q_T \frac{q_T^2}{\left(q_T^2 + \omega^2/\gamma^2\right)} F_N^2(q^2), \tag{4}$$

where $q^2 = (q_T^2 + \omega^2/\gamma^2)^2$ the 4-momentum transfer of the relativistic nuclei projectile, and $F_N(q^2)$ is the nuclear form factor of the equivalent photon source.

In the calculations below we use a simple approximation of the nuclear form factor. This approximation corresponds to an exponentially decreasing charge distribution of the nucleus, that the mean square radius is adjusted to fit the experimental value [36–38],

$$F_N(q^2) = \frac{\Lambda^2}{\Lambda^2 + q^2}, \quad \text{where} \quad \Lambda^2 = \frac{0.164\text{GeV}^2}{A^{2/3}}, \tag{5}$$

where $\Lambda = 0.091$ GeV for $^{197}$Au, $\Lambda = 0.088$ GeV for $^{208}$Pb, and $\Lambda = 0.065$ GeV for $^{238}$U.

In the semi-coherent two-photon interaction process at the ultra-peripheral nucleus-nucleus collisions, the momentum for photons are $q_1 = (\omega_1, \boldsymbol{q}_{1T}, q_{1z})$ and $q_2 = (\omega_2, \boldsymbol{q}_{2T}, q_{2z})$, the total transverse momentum of bound $(l^+l^-)$ state is $\boldsymbol{p}_T = \boldsymbol{q}_{1T} + \boldsymbol{q}_{2T} \approx \boldsymbol{q}_{1T}$, where $\boldsymbol{q}_{iT}$ is the transverse momentum of the $i$-th photon. Consequently, the differential cross section for the ultraperipheral nucleus-nucleus collisions can be written in the terms of bound $(l^+l^-)$ state transverse momentum as the following

$$\frac{d\sigma}{d^2p_T dy} = \frac{8Z^4\alpha^2}{\pi^2}(2J+1)\frac{\Gamma_{(l^+l^-)\to\gamma\gamma}}{M^3}\frac{F_N^2\left(p_T^2 + \frac{M^4}{16p_T^2\gamma^2}\right)}{p_T^2}$$

$$\times \int d^2q_T q_{2T}^2 \frac{F_N^2\left(q_{2T}^2 + \frac{M^4}{16p_T^2\gamma^2}\right)}{\left(q_{2T}^2 + \frac{M^4}{16p_T^2\gamma^2}\right)^2}, \tag{6}$$

where γ is the relativistic factor, and the transverse momentum of photon is $q_{2T} > 0.2$ GeV due to the single track acceptance condition [26].

In impact parameter space, the total number of photons from a relativistic nucleus can be obtained by integrating over all impact parameters larger than some minimum, which is typically given by the nuclear radius. In relativistic nucleus collisions, the impact parameter representation provides the best way to incorporate effects of strong absorption. This implies that hadronic interactions will dominate in relativistic nucleus collisions where both hadronic and electromagnetic interactions are possible. Considering the accurate hadronic interaction probabilities for the equivalent two-photon luminosity in the ultra-peripheral heavy ion collisions, the cross section of exotic electromagnetic bound systems $(l^+l^-)$ produced by coherent two-photon process can be expressed as

$$d\sigma = \hat{\sigma}_{\gamma\gamma \to (l^+l^-)}(W) dN_1(\omega_1, \vec{b}_1) dN_2(\omega_2, \vec{b}_2) S^2_{abs}(\vec{b})$$
$$= d\omega_1 d\omega_2 d^2 b_1 d^2 b_2 \hat{\sigma}_{\gamma\gamma \to (l^+l^-)}(W) \frac{dN_1(\omega_1, \vec{b}_1)}{d\omega_1 d^2 b_1} \frac{dN_2(\omega_2, \vec{b}_2)}{d\omega_2 d^2 b_2}, \quad (7)$$

where the photon spectrum can be expressed in terms of the charge form factor $F_N(q^2)$ as follows [39–42]

$$\frac{dN(\omega, \vec{b})}{d\omega d^2 b} = \frac{Z^2 \alpha}{\pi^2 \omega} \left[ \int_0^\infty dq_T q_T^2 \frac{F_N^2\left(q_T^2 + \omega^2/\gamma^2\right)}{q_T^2 + \omega^2/\gamma^2} J_1(bq_T) \right]^2, \quad (8)$$

here $J_1(x)$ is Bessel function.

The absorptive factor $S^2_{abs}(\vec{b})$, that excludes the overlap between the colliding nuclei and allows to take into account only ultraperipheral collisions, can be expressed in terms of the probability of interaction between the nuclei at a given impact parameter [29, 41],

$$S^2_{abs}(\vec{b}) = 1 - P_H(\vec{b}), \quad (9)$$

with

$$P_H(\vec{b}) = 1 - \exp[\sigma_{NN} T_{AA}]$$
$$= 1 - \exp\left[\sigma_{NN} \int d^2 r T_A(\vec{r}) T_A(\vec{r} - \vec{b})\right], \quad (10)$$

where $T_A$ is nuclear thickness function [43], and $\sigma_{NN}$ being the total hadronic interaction cross section, 52mb at RHIC and 88mb at LHC [44].

In the transformations $d\omega_1 d\omega_2 = (W/2)dWdy$, the differential cross section for the coherently two-photon interaction process can be written in the terms of bound $(l^+l^-)$ state rapidity as the following

$$\frac{d\sigma}{dy} = 8\pi^2 (2J+1) \frac{\Gamma_{(l^+l^-) \to \gamma\gamma}}{M^3} \int d^2 b_1 d^2 b_2 S^2_{abx}(\vec{b})$$
$$\times \frac{Z^2 \alpha}{\pi^2} \left[ \int_0^\infty dq_{1T} q_{1T}^2 \frac{F_N^2\left(q_{1T}^2 + \omega_1^2/\gamma^2\right)}{q_{1T}^2 + \omega_1^2/\gamma^2} J_1(b_1 q_{1T}) \right]^2 \quad (11)$$
$$\times \frac{Z^2 \alpha}{\pi^2} \left[ \int_0^\infty dq_{2T} q_{2T}^2 \frac{F_N^2\left(q_{2T}^2 + \omega_2^2/\gamma^2\right)}{q_{2T}^2 + \omega_2^2/\gamma^2} J_1(b_2 q_{2T}) \right]^2,$$

where $q_{iT}$ is the transverse momentum of the $i$-th photon.

III. NUMERICAL RESULTS

Since $Z\alpha \approx 0.6$ for relativistic heavy nuclei is large, the equivalent photon fluxes for the heavy nucleus become very large at the RHIC and LHC energies. Especially for the (semi-)coherently two-photon interactions, the photon flux is high enough that this interactions may be accompanied in the ultraperipheral nucleus-nucleus collisions the two-photon differential cross-section scales as $Z^4$.

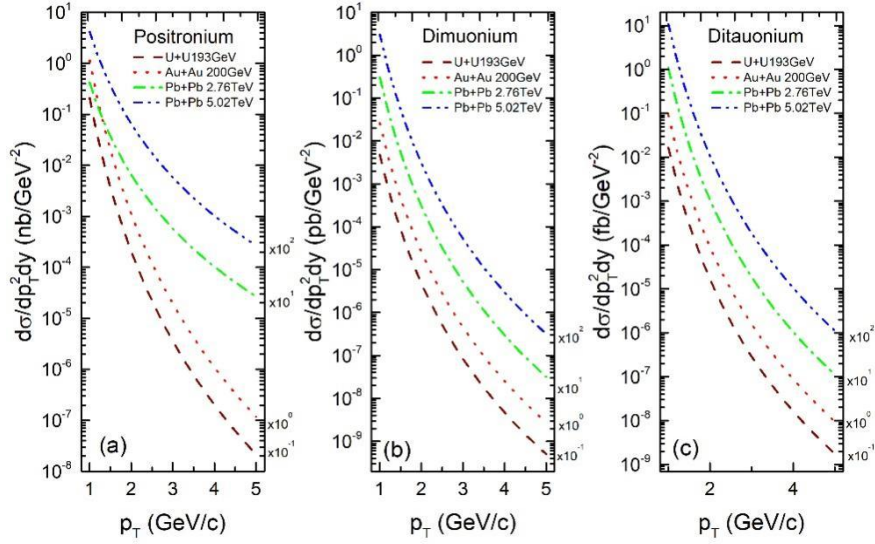

FIG. 1. The differential cross section for exotic electromagnetic bound systems ($l^+l^-$) production from the semicoherent two-photon interaction (without impact parameter b) in ultraperipheral heavy ion collisions at RHIC and LHC. The dashed line (wine line) is for U+U collisions with $\sqrt{s_{NN}} = 193 GeV$, the dotted line (red line) is for Au+Au collisions with $\sqrt{s_{NN}} = 200 GeV$, the dashed-dotted line (green line) for Pb+Pb collisions with $\sqrt{s_{NN}} = 2.76 TeV$, the dasheddotted-dotted line (blue line) is for Pb+Pb collisions with $\sqrt{s_{NN}} = 5.02 GeV$.

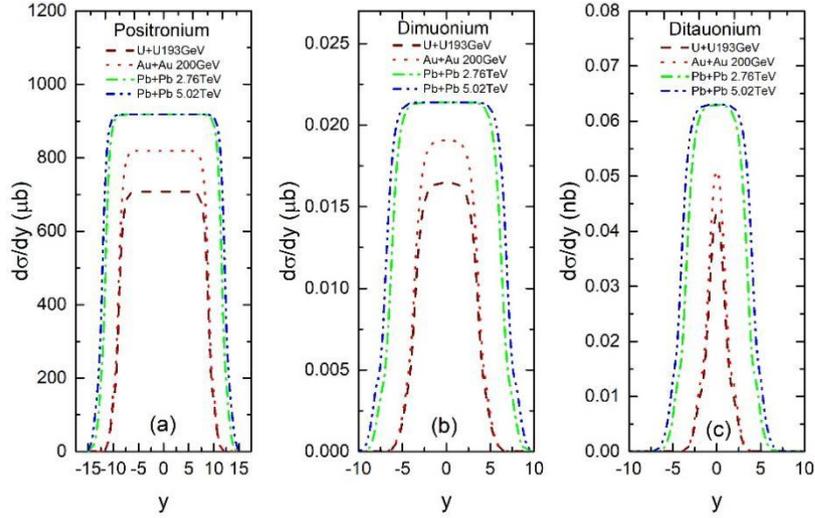

FIG. 2. The diffferential cross section for exotic electromagnetic bound systems ($l^+l^-$) production from the coherent two-photon interaction (with impact parameter *b*) in ultraperipheral heavy ion collisions at RHIC and LHC. The dashed line (wine line) is for U+U collisions with $\sqrt{s_{NN}} = 193 GeV$, the dotted line (red line) is for Au+Au collisions with $\sqrt{s_{NN}} = 200 GeV$, the dashed-dotted line (green line) for Pb+Pb collisions with $\sqrt{s_{NN}} = 2.76 GeV$, the dashed-dotted-dotted line (blue line) is for Pb+Pb collisions with $\sqrt{s_{NN}} = 5.02 GeV$.

Indeed, in the semi-coherently two-photon process, we keep the transverse momentum of one of the photons small, then the whole nucleus acts coherently. A charge distribution is necessary to the form factor, but spectra of positronium ($e^+e^-$), dimuonium ($\mu^+\mu^-$), and ditauonium ($\tau^+\tau^-$) under $\gamma\gamma \to (l^+l^-)$ interaction are depressed by the monopole form factor we used here. The differential cross section for large-$p_T$ exotic electromagnetic bound systems ($l^+l^-$) production at RHIC and LHC energies is potted in Fig. 1. the rapidity distribution of the exotic electromagnetic bound systems ($l^+l^-$) production by coherent two-photon interaction in the impact parameter space at RHIC and LHC are plotted in Fig. 2. The main sources of changes in the differential cross sections are the magnitude of the decay width and the mass of the produced particle, since mass difference between the positronium, dimuonium, and ditauonium is very large.

## IV. CONCLUSION

In summary, we have investigated the production of exotic electromagnetic bound systems ($l^+l^-$), such as positronium ($e^+e^-$), dimuonium ($\mu^+\mu^-$), and ditauonium ($\tau^+\tau^-$), from the semi-coherently coherently two-photon interaction process in equivalent photons approximation at RHIC and LHC energies. In the equivalent photon approximation, the effect of the electromagnetic field for the ultra-relativistic nucleus is replaced by the flux of photons. By using a charge distribution form factor, we show the transverse momentum distribution of exotic electromagnetic bound systems ($l^+l^-$) produced by semi-coherent two-photon interactions, as well as rapidity distribution of the exotic electromagnetic bound systems ($l^+l^-$) the coherent two-photon interactions with considering the effects of strong absorption in impact parameter space. Our calculations show that the large values of the differential cross sections for nucleus-nucleus collisions can be obtained with the semi-coherent and coherent approach in ultra-peripheral heavy ion collisions at the RHIC and LHC energies.

## V. ACKNOWLEDGEMENTS

This work is supported by Heilongjiang Science Foundation Project under Grant No. LH2021A009, and National Natural Science Foundation of China under Grant No. 12063006.